\documentclass[%
preprintnumbers,
 amsmath,amssymb,
 aps,
]{revtex4-2}

\usepackage{graphicx}
\usepackage{dcolumn}
\usepackage{float}
\usepackage{dblfloatfix}
\usepackage{bm}

\begin{document}

\preprint{APS/123-QED}

\title{Robustness Assessment of Complex Networks using the Idle Network}

\author{Marcus Engsig}
 \email{marcus.w.engsig@gmail.com}
\affiliation{%
 Department
of Science and Engineering, Sorbonne University Abu Dhabi, Abu Dhabi, United Arab Emirates.
}%

\author{Alejandro Tejedor}%
 \email{alej.tejedor@gmail.com}
\affiliation{%
 Department
of Science and Engineering, Sorbonne University Abu Dhabi, Abu Dhabi, United Arab Emirates.
}%

\affiliation{%
 Department of Civil and Environmental Engineering, University of California, Irvine, Irvine, CA 92697, USA
}%


\author{Yamir Moreno}
\email{yamir.moreno@gmail.com}
\affiliation{%
 Institute for Biocomputation and Physics of Complex Systems (BIFI), Universidad de Zaragoza, 50018 Zaragoza, Spain
}%
\affiliation{
 Departamento de F\'isica Te\'orica, Universidad de Zaragoza, 50009 Zaragoza, Spain
}%
\affiliation{Institute for Scientific Interchange, ISI Foundation, Turin, Italy
}%


\date{\today}

\begin{abstract}
Network robustness is an essential system property to sustain functionality in the face of failures or targeted attacks. Currently, only the connectivity of the nodes unaffected by an attack is utilized to assess robustness. We propose to incorporate the properties of the emerging connectivity of the nodes affected by the attack (Idle Network), which is demonstrated to contain pertinent information about network robustness, improving its assessment accuracy. The Idle network information offers the potential to generalize models, enabling them to estimate robustness for unseen attacks.

\end{abstract}

\maketitle


The representation of complex systems as networks, where system components are abstracted as nodes and their interactions as links, has allowed us to further our understanding of system structure and dynamics in fields as diverse as biology, engineering, economics, and geosciences \cite{Watts1998,Barabasi1999,Newman2010,Boccaletti2006,Barrat2008,Rodriguez-Iturbe1997,Bullmore2009, Kivela2014, Tejedor2017_DE}. Particularly, network theory has been instrumental in developing methodologies to assess the robustness of interconnected systems such as power grids, the internet, and airports, in the face of random failures or targeted attacks \cite{Sole2008, Cohen2000,Cohen2001,Wuellner2010,Schneider2011}.
The robustness of a network can be defined as the ability of the network to maintain functionality whilst undergoing an attack (sequential node removal). In a world where critical infrastructures and their connectivity are potential targets of malicious attacks, it is paramount to identify the key network properties that determine robustness for a given attack. Since the pioneering work by {\it Albert et al.} \cite{Albert2000}, a vast literature has  presented methodologies and metrics to quantify network robustness \cite{Albert2000,Cohen2000,Callaway2000,Cohen2001,Holme2002,Motter2002,Buldyrev2010,Min2014}. However, current methodologies to assess network robustness focus mainly on the connectivity of the nodes unaffected (Active Network) by the attack, while the connectivity of the affected nodes (Idle Network) has received minimal attention \cite{Tejedor2017_AI}. In this study, we demonstrate the benefit of including information about the Idle Network in assessing network robustness.

\begin{figure}[ht]
    \centering
    \includegraphics[width = 10 cm]{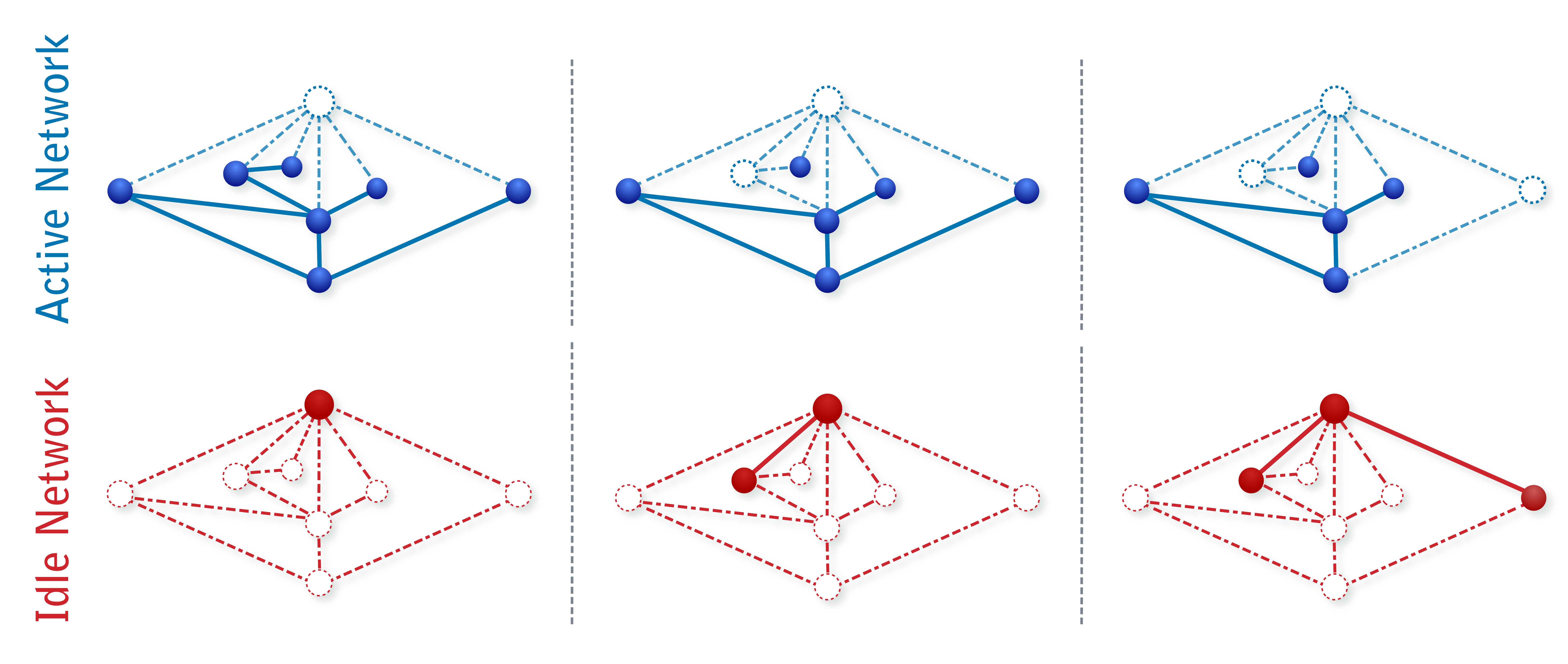}
    \caption{Illustration of the Active and Idle networks at different stages of an attack.}
    \label{fig:ActiveIdle}
\end{figure}

Let us formally define the \emph{Active} and \emph{Idle Networks}, which naturally emerge from an attack process acting on a network \cite{Tejedor2017_AI}. Attacking a network is synonymous to a process of \emph{sequential node removal}. Consider an initial network $\mathcal{N}$ that consists of $N$ nodes, denoted $\{n_i\}: i = 1,... N$, connected by a set of links $\{(n_i, n_j)\}$. The sequential node removal process starts at $t=0$ with the original network $\mathcal{N}$, and an attack strategy $D$, that is a function of the properties of $\mathcal{N}$. For every discrete time step $t>0$, the attack eliminates a chosen node $n_i$ and all its corresponding links $(n_i, \cdot )$, resulting in a new network, formed by the set of nodes and links that is unaffected by the attack; we denote this the Active Network $\mathcal{N}_A (t)$. The attack process also gives rise to the Idle Network $\mathcal{N}_I (t)$, which consists of the entire set of nodes removed from the Network $\mathcal{N}$ up to time $t$, and the links originally existing among them (see Fig. \ref{fig:ActiveIdle}). We can mathematically express a given attack strategy $D$ acting on a network $\mathcal{N}$, as the decomposition of $\mathcal{N}$ into the Active $\mathcal{N}_A (t)$ and Idle $\mathcal{N}_I (t)$ networks. 
\begin{equation}
    D: N\to \{N_A(t), N_I(t)\}, \quad t = 1,...N
    \label{attack}
\end{equation}
It is clear that with respect to the nodes, the Active and Idle networks are complementary, implying that the union of the nodes in $\mathcal{N}_A(t)$ and $\mathcal{N}_I (t)$ is the set of nodes in $\mathcal{N}$. However, this is not the case for the connectivity of the nodes, as it is neither complementary nor symmetric. 
When a node is removed, all its links are removed from the Active Network. Yet, from the set of links removed by the attack, only the subset that connects affected (removed) nodes is included in the Idle network. We argue that the information about the connectivity  of the affected nodes by an attack, which is just available in the Idle Network, provides important information on the effectiveness of the attack and, therefore, on the robustness of the attacked network. Thus, our research hypothesis can be stated as follows: \emph{there exists non-redundant information on the robustness of a network undergoing an attack in the Idle Network structure.}  

To test this hypothesis, we will extract indicators from the Active and Idle Networks to benchmark our capacity to assess network robustness using only Active indicators (traditional approach) versus incorporating Idle indicators as well. Particularly, we choose two simple indicators to model robustness: (i) The \textit{largest cluster size} $C$, defined as the ratio of the number of nodes in the largest cluster (set of connected nodes) over the number of nodes $N$ in the initial network, quantifies the effect of the attack in breaking down (building up) the Active (Idle) network in terms of its size. Note that this metric does not encompass the effectiveness of the connectivity of these networks. (ii) The \textit{link fraction} $L$ is the number of links in the Active (Idle) network, normalized by the total number of links in the initial network $\mathcal{N}$. This indicator describes how the attack removes (adds) links and thus provides information about how well-connected  the nodes are in the Active (Idle) Network. Both of these indicators are normalized to be between $[0,1]$, and are monotonically decreasing (increasing). These indicators were chosen such that, in complement, they have information on the overall functionality of the network and therefore, on its robustness.

Following previous studies \cite{Trajanovski2013,Ventresca2015,Williams2016,Cats2020}, we utilize the \textit{efficiency}, $E$, as a proxy for robustness. Recall that $E$ of a network $\mathcal{N}$ with $N$ nodes is defined as the standardized sum of the reciprocal of the shortest paths $d_{i,j}$ between all pair of nodes $i$ and $j$; 
\begin{equation}
    E = \frac{1}{N(N-1)}\sum_{i,j \in \mathcal{N}, i \neq j} \frac{1}{d_{i,j}}
    \label{eq:geodesicmean}
\end{equation}
Note that, if two nodes $i,j$ are disconnected, then $\frac{1}{d_{i,j}} = 0$, as the distance between the two nodes is infinite.  $E$ is normalized to always start at $1$, by dividing all values of $E$ for a single evolution by the value of the efficiency for the intact network. We underline that $E$ is a property of the Active Network, as it is solely a function of the \textit{adjacency matrix} of the Active Network.  

Given the two indicators and the proxy for robustness, we transform our hypothesis into a regression problem. Thus, we evaluate the difference in estimation accuracy achieved via a neural network when only Active indicators are included in the training set, and when Idle indicators are also included. More specifically, we use a forward-feeding and back-propagating artificial neural network with $3$ hidden layers of $10$ neurons per layer, each with \texttt{ReLu} activation functions; set to optimize validation squared residual loss. Each neural network was implemented with a dataset of $200$ attack sequences, with a $\frac34$ train, $\frac18$ test, and $\frac18$ validation split. The output of the neural network is the estimation of the efficiency as the proxy for robustness. In order to verify our hypothesis, the estimation accuracy must increase when the neural network is granted the Active and Idle indicators, compared to the estimation produced using the Active indicators alone. 

\begin{figure*}[!ht]
    \centering
    \includegraphics[width=\textwidth]{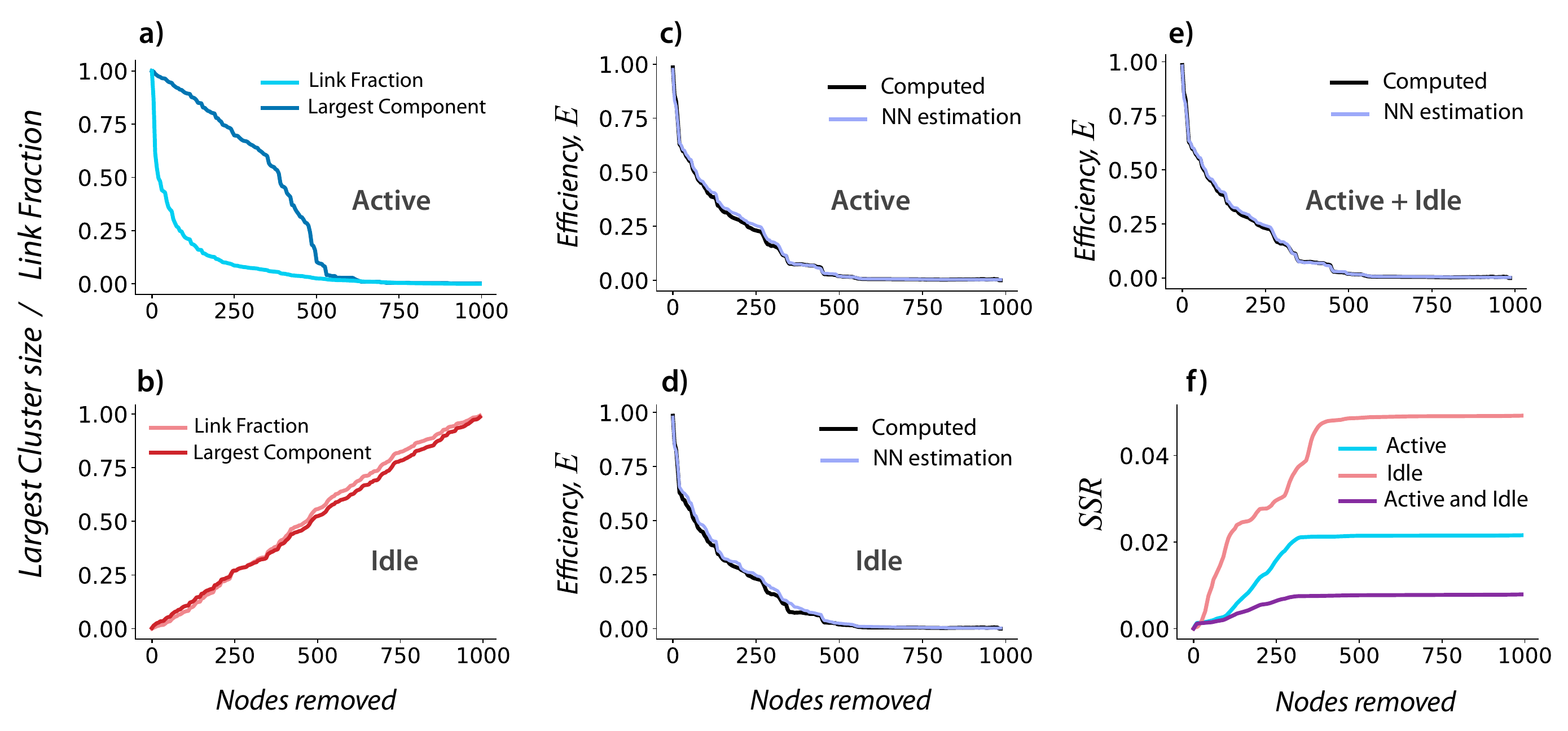}. 
    \caption{Evolution of the (a) Active and  (b) Idle indicators (largest component and link fraction) for a scale-free network with $N=1000$ nodes ($\bar{k} = 12$), undergoing a degree attack. Estimation of the network Efficiency as a function of the attack stage via a Neural Network using indicators from (c) only Active or (d) only Idle Networks. (e) Estimation of the efficiency at different stages of the attack via a neural network trained with both the Active and the Idle indicators. (f) The the sum of the squared residuals ($SSR$) computed for the different estimations as the function of the attack stage is displayed.}
    \label{fig:specific}
\end{figure*}

Our study investigates different stochastically generated synthetic network topologies and attack strategies to test our hypothesis systematically. Namely, we test the robustness estimation for random (Erdos Renyi \cite{Erdos1984}), scale-free (using a configuration model \cite{Catanzaro2005}), and small world (Strogatz-Watts \cite{Watts1998}) topologies, undergoing three different attack strategies: targeted (degree), random failure, and random spreading \cite{Tejedor2017_AI}. Furthermore, the different topologies were explored for varying initial link densities, as characterized by $\bar{k}$ (average degree of the initial network $\mathcal{N}$). The tested link densities for all of the synthetic topologies correspond to $\bar{k}\in \{3,6,12,24\}$. Thus, we have explored $36$ combinations of topologies, attacks, and link densities. For each of these combinations, $200$ different stochastic topologies were generated and exposed to a full attack evolution, where the indicators and efficiency were calculated at the different stages of the attack (see \textit{Supplemental Material (SM)}). 

Fig. \ref{fig:specific} displays a representative case to illustrate our results. As expected, the estimation of network robustness using Active indicators (Fig. \ref{fig:specific}a) is quite accurate (Fig. \ref{fig:specific}c - $SSR = 0.02 \pm 0.01$). Noticeably, the Idle indicators alone (Fig. \ref{fig:specific}b) allowed us to estimate fairly well ($SSR = 0.06 \pm 0.05$) the trend of the evolution of the efficiency as shown in Fig. \ref{fig:specific}d. Most importantly, as hypothesized, by combining the indicators of the Active and Idle networks, we obtained a more accurate estimation of network robustness ($SSR = 0.01 \pm 0.006$) (Fig. \ref{fig:specific}e). Additionally, the increased accuracy in the estimation is consistent throughout all stages of the attack (Fig. \ref{fig:specific}f).  Our results for the whole data set of network topologies and attacks  demonstrate systematically that Active indicators, when combined with Idle indicators, increase the accuracy in the estimation of  robustness from $20\%$ to $900\%$ depending on the topology and attack, verifying our hypothesis (see SM). 

Guided by the following observation made from the data set of topologies and attacks analyzed in this work (see SM): ``the more complex (i.e., more variable at different scales) the efficiency curves are, the higher the improvement in the accuracy of robustness assessment by acknowledging Idle indicators'', we  investigate the potential role of Idle information in distilling variability in the data set to improve network robustness estimation.  To this goal, we systematically explore the effect of variability in the training set in estimating robustness. More specifically, we trained neural networks with training sets with increasing variability by combining different topologies, attacks, and link densities (including a data set consisting of all combinations), and we compared the estimation accuracy when only Active indicators are considered, and when Active and Idle indicators are both included.

\begin{figure*}[t]
\includegraphics[width = \textwidth]{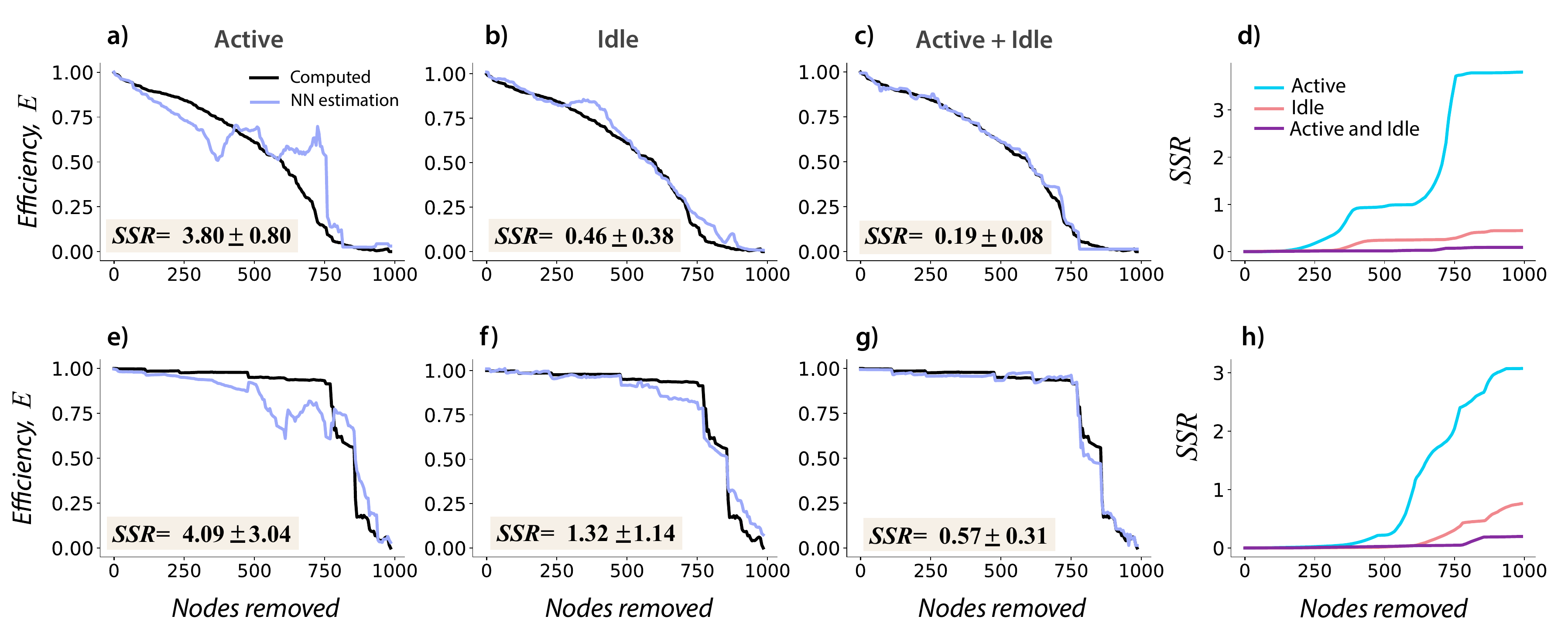}
\caption{\label{fig:generalized} Performance of a neural network in estimating network robustness when trained with the entire combined data set of topologies, attacks, and link densities. Results for three different models are presented: neural network trained with (a,e)  Active indicators, (b,f) Idle indicators, and (c,g) both Active and Idle indicators. The estimations of these models is displayed for a Erdos Renyi topology with $\bar{k} = 6$ undergoing degree targeted attacks (top panels), and for scale free networks of $\bar{k} = 12$ undergoing random attacks (bottom panels). The mean and standard deviation of the sum of the squared residuals ($SSR$) computed for a testing set, consisting of 25 synthetic topologies undergoing their respective attacks, is also reported as an inset in the respective panels. Cumulative values of $SSR$ as a function of the attack stage for the (d) Erdos Renyi and (h) sale free topologies are also displayed.}
\end{figure*}

Fig. \ref{fig:generalized} shows the model outputs for the most generalized case:  data for all the three topologies, four densities, and three attacks are included in the training set. The results are apparent, the inclusion of Idle indicators (see \ref{fig:generalized} c and g) produce exceedingly good predictions when compared with those achieved via only Active indicators (see \ref{fig:generalized} a and b). When the difference between the model output and the true value of robustness ($SSR$) is computed as a function of the attack stage (see \ref{fig:generalized} d and h), a consistent pattern is observed: Active and Idle indicators combined outperformed the Active indicators alone during the most significant part of the attack sequence. 



As expected, a  general trend is also observed (see SM): the more heterogeneous the training set is, the less accurate is the estimation of network robustness done by all three neural networks  (trained with: Active indicators only, Idle indicators only, and Active and Idle indicators). However, the rate of performance deterioration is not similar, in fact, it is not comparable. As soon as variability is introduced in the training set, the neural network using the Active indicators exclusively is not able to estimate even the general trend, let alone the variability. Whereas the neural network trained using both the Active and Idle indicators is able to estimate the general trend very well and a majority of the variability. This trend is consistent for all of the topologies tested (see Fig. \ref{fig:generalized} and SM). 

Two further remarks are noteworthy from this part of the work:  (i) In a surprising number of times,  the robustness estimations obtained via the neural networks trained exclusively with Idle indicators are significantly more accurate than those produced by the neural networks trained only with Active indicators, highlighting the non-redundant relevant information content in the Idle network. (ii) In select cases, the  Neural Network trained with all topologies, densities, and attacks outperforms in terms of accuracy the estimation of robustness made by a Neural Network trained purely for a specific topology, attack, and link density, highlighting the value of Idle indicators in interpreting the overall variability in the dataset to improve estimations for specific cases. Therefore, we claim that the intrinsic information in the Idle Network, jointly with the Active indicators, allow the neural networks to \textit{navigate} the variability in the training set to maintain an enhanced accuracy in assessing network robustness.

\begin{figure}[t]
\includegraphics[width = 0.5\linewidth]{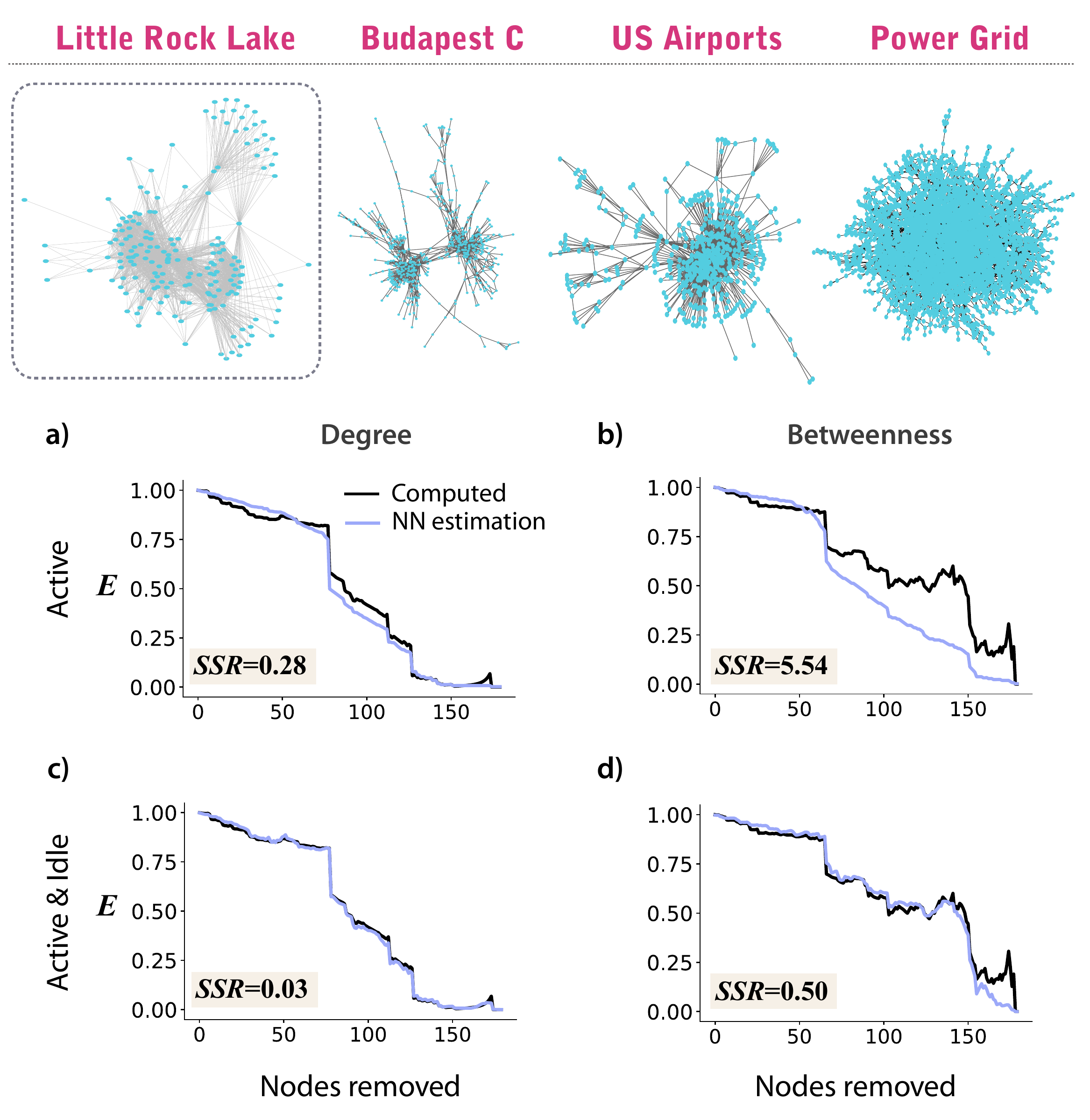}
\caption{\label{fig:realnetworks} Illustration of the four real network topologies tested (top panels), along with the results for the Little Rock Lake. An artificial neural network for the Little Rock Lake was trained by attacking the topology with $200$ full stochastic degree evolution's, and attempts to predict a previously unseen attack scheme (stochastic betweenness attack). In these attack schemes, the probability of removing a specific node is proportional to the degree or betweenness centrality. The sum of squared residuals is displayed for a single evolution.}
\end{figure}

Our previous results have clearly demonstrated that the Idle network contains relevant information, useful for improving the assessment of network robustness. However, the degree of improvement in that assessment varies depending on the attack and network topology. Acknowledging that the used synthetic networks lack some properties often exhibited by  real-world networks (e.g., modularity), here, we further test the relevance of Idle network information in assessing robustness of real networks. To do that, we simulate stochastic targeted degree attacks on real-world topologies, where the probability of removing a given node is proportional to its original degree. We also evaluate the role of Idle information in generalizing the estimation robustness for an unseen attack (e.g., based on betweenness centrality). Particularly, we first train a neural network using only Active indicators resulting from 200 node removal sequences obtained by following a stochastic targeted degree attack strategy. Our results show a fairly good estimation of our proxy of robustness (See Fig. \ref{fig:realnetworks}a -Little Rock Lake Food Web \cite{Martinez1991}). However, suppose that trained neural network is used to estimate the network robustness of the same network topology under a stochastic targeted betweenness attack (unseen attack). In that case, the estimation fails to reproduce the evolution of the true value during the vast majority of the attack sequence (see Fig. \ref{fig:realnetworks}c). On the other hand, if a neural network is trained with the Active and Idle indicators of the same 200 node removal sequences (stochastic targeted degree attacks), not only we obtain better accuracy in estimating network robustness under stochastic targeted degree attacks (see Fig. \ref{fig:realnetworks}b -Little Rock Lake Food Web), but also that neural network provides an exceptionally well-maintained accuracy in the estimation of network robustness for a previously unseen attack (stochastic targeted betweenness attack) for the vast majority (and relevant) part of the attack sequence (See Fig. \ref{fig:realnetworks}d -Little Rock Lake Food Web). These results have been tested for several real-world networks (Little Rock Lake Food Web \cite{Martinez1991}, Budapest Connectome \cite{Szalkai2015}, and US airports \cite{Colizza2007} - See SM), corroborating our two previous findings, namely, (i) Idle network information systematically improves our capacity to estimate network robustness, and (ii) Idle information allows us to retain accuracy in network robustness estimation under scenarios of enhanced variability, both in the training set and out-of-sample (e.g., altered attack strategies).

Our results indicate that the key role of Idle indicators is to partially harness the existing information in the internal variability of the training set to gain estimation power (i) in the face of variability in the training set (either from its intrinsic stochastic variability or due to the inclusion of different topologies and attacks in the training set), and (ii) for unforeseen attacks and topological features that generate variability compatible with that observed in the training set. Thus, the Idle network information is instrumental for our model (neural network) to interpret variability and improve the robustness assessment.  However, if the variability in the data set is minimal (e.g., targeted attack in a sparse scale-free network), the gain achieved by including Idle indicators would be incremental. Furthermore, the indicators chosen in this study (size of the largest cluster and link fraction) could be particularly clumsy in encoding complementary information on network robustness to that encoded by the Active indicators for certain network topologies (e.g., spatial networks such as the power grid \cite{Watts1998}), and therefore, these Idle indicators might be ineffective in enhancing network robustness assessment in those cases.

We want to finally remark that this study uses a neural network as a tool to turn our hypothesis into a regression problem. The chosen neural network architecture and typology to estimate our proxy of robustness is not intended to be optimal, but to demonstrate the information content and role of the Idle network in the assessment of network robustness. Thus, for example, we anticipate that using convolutional neural networks may improve the accuracy of robustness estimation. Such further improvements in the accuracy of estimating efficiency can lead to important implications of our work, since neural networks trained for generalized data sets would offer a light way to estimate network efficiency, which otherwise is a computationally very demanding quantity to be calculated. 


Assessing network robustness accurately is essential to ensure the correct and sustained functionality of many natural and engineered systems. Our study shows that there is non-redundant and pertinent information on the robustness of a network in the so-called Idle network. The inclusion of Idle information in models to assess network robustness allows us to improve the accuracy of our estimations for a specific network topology and attack and equips models with the capability to interpret in-sample and out-sample variability to preserve estimation power amid noise and unseen variability. Thus, evaluating network robustness in the light of the Idle Network constitutes a conceptual paradigm shift that could improve the quality and accuracy of its assessment and might lead to new strategies to guide enhanced network resilience.   




\begin{acknowledgments}
A.T. acknowledges financial support support from the NSF Earth Sciences Directorate Grant EAR-1811909. Y.M. acknowledges support by the Government of Aragón and “ERDF A way of making Europe” funds through grant E36-20R, by Ministerio de Ciencia e Innovación, Agencia Española de Investigación (MCIN/AEI/10.13039/501100011033) through grant PID2020-115800GB-I00, and by Soremartec S.A. and Soremartec Italia, Ferrero Group. The funders had no role in study design, data collection, and analysis, decision to publish, or preparation of the manuscript.

\end{acknowledgments}

\bibliography{Biblio_Nets}










\newpage

\begin{center}

  \vspace*{-2cm}\makebox[\textwidth]{\includegraphics[page=1,width=\paperwidth]{{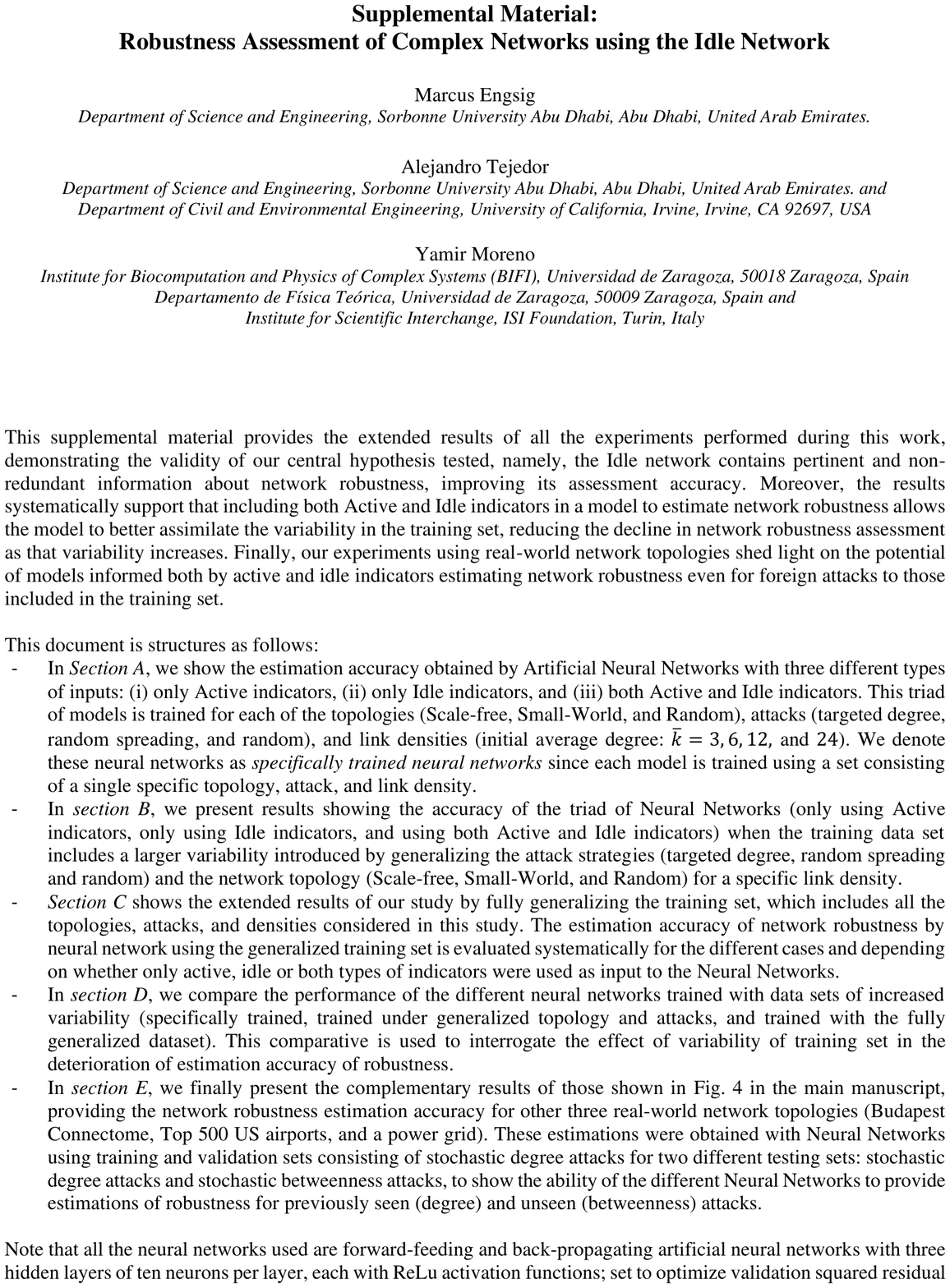}}}
  \newpage
   \vspace*{-2cm}\makebox[\textwidth]{\includegraphics[page=2,width=\paperwidth]{{Engsig_etal22_SM.pdf}}}
     \newpage
   \vspace*{-2cm}\makebox[\textwidth]{\includegraphics[page=3,width=\paperwidth]{{Engsig_etal22_SM.pdf}}}
     \newpage
   \vspace*{-2cm}\makebox[\textwidth]{\includegraphics[page=4,width=\paperwidth]{{Engsig_etal22_SM.pdf}}}
     \newpage
   \vspace*{-2cm}\makebox[\textwidth]{\includegraphics[page=5,width=\paperwidth]{{Engsig_etal22_SM.pdf}}}
     \newpage
   \vspace*{-2cm}\makebox[\textwidth]{\includegraphics[page=6,width=\paperwidth]{{Engsig_etal22_SM.pdf}}}
     \newpage
   \vspace*{-2cm}\makebox[\textwidth]{\includegraphics[page=7,width=\paperwidth]{{Engsig_etal22_SM.pdf}}}
     \newpage
   \vspace*{-2cm}\makebox[\textwidth]{\includegraphics[page=8,width=\paperwidth]{{Engsig_etal22_SM.pdf}}}
     \newpage
   \vspace*{-2cm}\makebox[\textwidth]{\includegraphics[page=9,width=\paperwidth]{{Engsig_etal22_SM.pdf}}}
     \newpage
   \vspace*{-2cm}\makebox[\textwidth]{\includegraphics[page=10,width=\paperwidth]{{Engsig_etal22_SM.pdf}}}

\end{center}

\end{document}